\documentclass[aps,pre,10pt,superscriptaddress,twocolumn,amsmath,amssymb,final,letterpaper]{revtex4-1}
\usepackage{graphicx}
\usepackage{bm}
\usepackage[utf8]{inputenc}
\usepackage[raggedright]{subfigure}
\usepackage[colorlinks=true,linkcolor=blue,urlcolor=blue,citecolor=blue,anchorcolor=blue]{hyperref}
\hypersetup{pdfdisplaydoctitle=true,pdftitle={Construction of and efficient sampling from the simplicial configuration model}}
\hypersetup{pdfauthor={J.-G. Young, G. Petri, F. Vaccarino, A. Patania}}
\hypersetup{pdfsubject={Construction of and efficient sampling from the simplicial configuration model}}

\newcommand{\avg}[1]{\langle #1 \rangle}

\usepackage[usenames,dvipsnames]{xcolor}

\begin{document}
\title{Construction of and efficient sampling from the simplicial configuration model}
\date{\today}

\author{Jean-Gabriel Young}
\email{jean-gabriel.young.1@ulaval.ca}
\affiliation{D\'epartement de Physique, de G\'enie Physique, et d'Optique, Universit\'e Laval, Qu\'ebec (Qu{\'e}bec), Canada}
\author{Giovanni Petri}
\affiliation{ISI Foundation, Torino, Italy}
\author{Francesco Vaccarino}
\affiliation{ISI Foundation, Torino, Italy}
\affiliation{Dipartimento di Scienze Matematiche, Politecnico di Torino, Torino, Italy}
\author{Alice Patania}
\email{alice.patania@isi.it}
\affiliation{ISI Foundation, Torino, Italy}
\affiliation{Dipartimento di Scienze Matematiche, Politecnico di Torino, Torino, Italy}

\begin{abstract}
Simplicial complexes are now a popular alternative to networks when it comes to describing the structure of complex systems, primarily because they encode multinode interactions explicitly.
With this new description comes the need for principled null models that allow for easy comparison with empirical data.
We propose a natural candidate, the \emph{simplicial configuration model}.
The core of our contribution is an efficient and uniform Markov chain Monte Carlo sampler for this model.
We demonstrate its usefulness in a short case study by investigating the topology of three real systems and their randomized counterparts (using their Betti numbers).
For two out of three systems, the model allows us to reject the hypothesis that there is no organization beyond the local scale.
\end{abstract}
\maketitle

Network science's approach to complexity rests onto the tacit hypothesis that the structure of complex systems is reducible to the pairwise interaction of their constituents.
It is often a valid premise and, as a result, network science has been extremely successful in, e.g., both predicting \cite{pastor2015epidemic} and controlling \cite{liu2016control} the behavior of complex systems, inferring their function from their structure \cite{porter2009communities,newman2012communities}, and so on.
Networks, however, might not be as ubiquitous as previously thought.
It has been shown recently that the structure of a number of complex systems, such as the brain \cite{dabaghian2012topological,giusti2016two}, protein interactions \cite{xia2014persistent} and social systems \cite{hebert2015complex,stolz2016topological}, cannot always be reduced to the sum of pairwise interactions. 
For these systems, it is now known that network representations can give an incomplete picture: When many-body interactions are broken down into multiple pairwise interactions (cliques), high-order information simply disappears \cite{zuev2015exponential}.

Simplicial complexes generalize graphs by encoding many-body interactions explicitly; they have hence been proposed as a complementary description of the structure of complex systems \cite{klein2014,curto2008cell,horak2009persistent,patania2017topological}.
Different from hypergraphs, they are equipped with an implicit notion of containment.
If nodes $(v_1,\hdots ,v_{q+1})$ are involved in a $q$-dimensional interaction, then it is implicit that all possible lower dimension interactions involving the same nodes also exist [for example $(v_1,\hdots v_{q})$ and $(v_1,v_3)$].
While it might appear constraining, this property actually arises in all systems where interactions are maximal, e.g., in scientific collaborations (largest cohesive group of collaborators) or gene activation pathways (largest group of collectively activated genes).
Furthermore, it is found in many processed relational datasets, e.g., in \textit{clique complexes}, obtained by mapping the cliques of networks to simplices \cite{petri2013topological,sizemore2016classification}, or in filtered simplicial complexes \cite{horak2009persistent}.
Simplicial complexes thus offer a natural and compact description of the structure of complex systems, both when-high order structures are explicitly available, or when they are extracted from low-order information.

This application of simplicial complexes has led to promising discoveries: We now better understand, for instance, how to detect large viral recombination events \cite{chan2013topology},  how brain networks reorganize under drugs \cite{petri2014homological}, and how the atomic structure of amorphous solids is hierarchically organized \cite{hiraoka2016hierarchical}. 
It has become crucial to establish the statistical significance of these findings, a task for which random null models will be  needed.
There is already a rich and growing literature on random simplicial complexes and topology, ranging from simplicial generalization of Erd\"{o}s-R\'enyi models, amendable to analytical treatment \cite{kahle2014topology,costa2016random}, to equilibrium formulations of simplicial complex ensembles \cite{courtney2016generalized,zuev2015exponential}, and growth models that reproduce various emergent patterns observed in real systems \cite{bianconi2016network,wu2015emergent}.
However, null models---in the sense of network science---are still wanting \cite{fosdick2016configuring,orsini2015quantifying}.

We address this issue by refining a recently proposed generalization \cite{courtney2016generalized} of the (simple) configuration model of network science \cite{molloy1995critical,newman2001random,fosdick2016configuring}, which we dub the simplicial configuration model (SCM).
Different from Ref.~\cite{courtney2016generalized}, we think of our model as a null hypothesis for real systems; we therefore develop a numerical and statistical toolbox instead of focusing on closed ensemble averages.
This entails a number of interesting results: 
One, we define the first simplicial configuration model able to describe arbitrary complexes, in line with our goal of obtaining a generic null model (Sec.~\ref{section:model}).
Two, we propose and analyze an efficient and rigorous sampling algorithm for this model (Sec.~\ref{section:sampler}).
Three, we use the model to investigate real datasets and show---now using sound statistical arguments---that the local structure of these systems does not always explain their mesoscale structure (Sec.~\ref{section:null_model}). 
We conclude by listing a few important open problems.

\section{Simplicial configuration model}
\label{section:model}
Informally, a labeled simplicial complex $K$ is the high-order generalization of a network.
Formally, it is a collection of simplices incident on a node set $V=\{v_1,\hdots ,v_n\}$ \cite{hatcher2000algebraic}.
A $q$--dimensional simplex---the generalization of an edge---is a tuple of $q+1$ distinct nodes $(v_1,\hdots ,v_{q+1})$; we say that this simplex is incident on $v_1,\hdots ,v_{q+1}$.
All simplices not included in a larger simplex are called the \emph{facets} of the complex, whereas a contained simplex is called a \emph{face}; e.g., if $K$ comprises of $\sigma=(v_1,v_2)$ and $\tau=(v_1,v_2,v_3)$, then $\sigma$ is a face of facet $\tau$.
It is always assumed that if facet $\sigma = (v_i,\hdots ,v_j)$ is in the simplicial complex $K$, all elements in the power set of $\sigma$ are also in $K$.
Therefore, faces need not be enumerated: The structure of a simplicial complex is fully specified by the list of its facets.

Departing from other recent contributions \cite{courtney2016generalized}, we define the degree $d_i$ of a node $v_i$ as the number of facets incident on $v_i$
and the size $s_i$ of a facet $\sigma_i$ as the number of nodes it contains (its \emph{dimension} plus one).
This local information  is summarized by the sequences $\bm{d}=(d_1,\hdots ,d_n)$ and $\bm{s}=(s_1,\hdots ,s_f)$, where $n$ is the number of nodes and $f$ is the number of facets.

With these notions in hand, we define the simplicial configuration model (SCM) as the uniform distribution over all labeled simplicial complexes with degree sequence $\bm{d}$ and facet size sequence $\bm{s}$.
In other words, if $\Omega(\bm{d},\bm{s})$ is the set of all labeled simplicial complexes with joint sequences $(\bm{d},\bm{s})$, then the SCM places a probability 
\begin{equation} \label{eq:scm_prob}
\Pr(K;\bm{d},\bm{s})=1/|\Omega(\bm{d},\bm{s})|
\end{equation}
on $K$ if it has sequences $(\bm{d},\bm{s})$, and 0 otherwise.
The model of Ref.~\cite{courtney2016generalized} is recovered by setting the size of all facets to a constant $s$.

This particular choice of definition for the SCM is natural for three reasons.
First, the SCM directly generalizes the simple CM of network science \cite{fosdick2016configuring};
when $s_i=\{1,2\}$ for all facets, one recovers a graph ensemble with degree sequence $\bm{d}$.
Second, the SCM does not include any correlation---the structure is maximally random beyond the local level.
This is reminiscent of the equivalent network model.
Third, the SCM can describe the local structure of any simplicial complex, since it allows for arbitrary degree and size sequences.
This property is \emph{not} common to all random models of simplicial complexes, for good reasons;
many models are constructed with a focus on the calculation of closed-form expression for a few properties (e.g., the asymptotic entropy) \cite{zuev2015exponential,courtney2016generalized,kahle2014topology}.
This commends simplifying assumptions, e.g., a regular facet size sequences \cite{courtney2016generalized}.
Our definition of the SCM forgoes these simplifications to accommodate arbitrary local structures, at the expense of analytical tractability.

\section{Efficient sampling algorithm}
\label{section:sampler}
\subsection{Constraints on the support}
For the SCM to be of any use, one needs to be able to sample from it.
This is far from a trivial problem, because there are numerous constraints on the support of the model.
It will be easier to see these constraints by first switching to the equivalent \emph{graphical} representation of simplicial complexes.

\begin{figure}
\includegraphics[width=\linewidth]{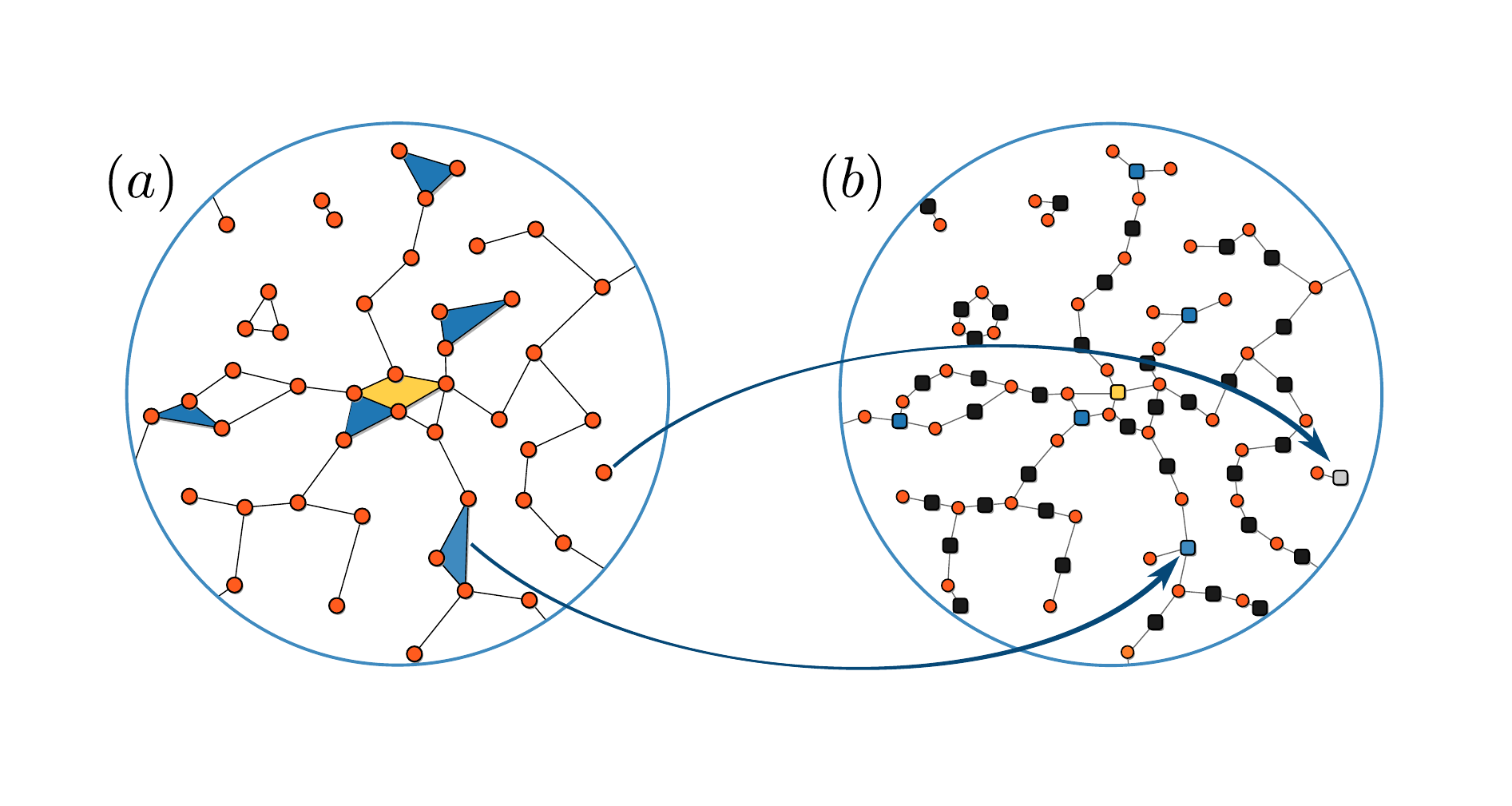}
\caption{ (a) Simplicial complex $K$ and (b) its graphical representation $B$.
In the bipartite graph, small square nodes represent facets and large orange  nodes represent the nodes of $K$. An edge connects a facet $\sigma_i$ and a node $v_j$ in $B$ if the facet $\sigma_i$ is incident on node $v_j$ in $K$.
Notice how some cliques are not filed (i.e., $k$ fully connected nodes do not necessarily form a size $k$), and how isolated nodes are attached to facets of size 1.
}
\label{fig:bipartite_graph}
\end{figure}

In this representation, facets are replaced by nodes (we denote by $F$ this new node set, and by $V\cup F$ the complete node set), and an edge connects facet $\sigma_i\in F$ to node $v_j\in V$ in $B$ if and only if $\sigma_i$ is incident to $v_j$ in $K$ (see Fig.~\ref{fig:bipartite_graph}).
Because $B$ encodes the structure of $K$ without ambiguity, we can think of the model in terms of either representations.

\begin{figure}[b]
\includegraphics[width=\linewidth]{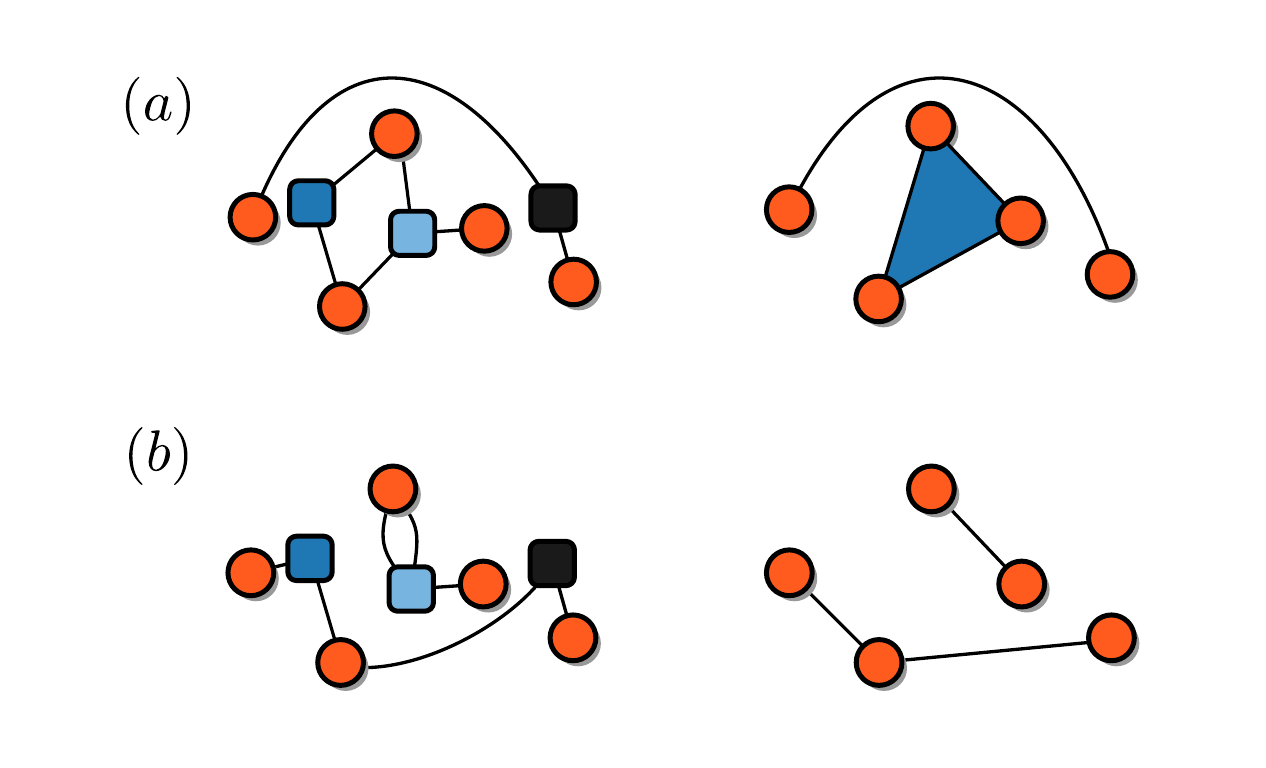}
\caption{Example of non-degree-preserving bipartite graphs. 
The two bipartite graphs (left column) encode the joint degree sequences $(\bm{d},\bm{s})= ([2,2,1,1,1], [3,2,2])$,  but the associated simplicial complexes (right column) have different size and degree sequences, respectively $(\bm{d}',\bm{s}')=([1,1,1,1,1], [3,2])$ and $(\bm{d}',\bm{s}')=([2,1,1,1,1], [2,2,2])$.
The disparities are due to the presence of (a) a fully included neighborhood, and (b) pairs of nodes connected by more than one edge.
}
\label{fig:degree-preserving}
\end{figure}

As such, one could be tempted to assume that sampling from the SCM of parameters ($\bm{d},\bm{s}$) is equivalent to uniformly sampling from all bipartite graphs with these degree sequences---a solved problem \cite{miklos2013towards}.
But this would be wrong: The mapping is not bijective.
This is, in fact, where the constraints on the support of the SCM become apparent \cite{courtney2016generalized}.
Let us introduce the notion of sequence preserving bipartite graph to formalize these constraints.
We say that a bipartite graph $B$ with joint degree sequences $(\bm{d},\bm{s})$ is sequence preserving if, upon interpretation of $B$ as a simplicial complex, one obtains a simplicial complex with facet size sequence $\bm{s}$ and generalized degree sequence $\bm{d}$.

Not all bipartite graphs are sequence preserving, and there are two reasons for this, both related to the fact that we think of the nodes in $F$ as facets.
The first reason is the inclusion of at least one facet: If there is a $\sigma_i\in F$ such that the neighborhood of $\sigma_i\subseteq\sigma_j$ for some $j\neq i$, then $B$ is not sequence preserving  [see Fig.~\ref{fig:degree-preserving}~(a)].
When this occurs, $\sigma_i$ is included in $\sigma_j$; the corresponding simplicial complex is thus either ill specified (facets cannot contain other facets, by definition) or does not have the same degree and size sequences as $B$ (if we simply remove $\sigma_i$).
For similar reasons, if two or more edges connect the same pair of nodes in $B$, then the graph is not sequence preserving [see Fig.~\ref{fig:degree-preserving}~(b)].

The sampling space would not be too constrained if these non-sequence-preserving bipartite graphs were rare.
Sampling would then be easy.
Unfortunately, it is straightforward to show that-non sequence-preserving graphs are   far more common than sequence--preserving ones, by adapting the calculations of Ref.~\cite{bender1978asymptotic}.
We find that the fraction $\phi$ of bipartite graphs with degrees $(\bm{d},\bm{s})$ not featuring parallel edges rapidly tends to
\begin{equation}
    \phi =e^{-\frac{1}{2}\left(\avg{d^2}/\avg{d}-1\right)\left(\avg{s^2}/\avg{s}-1\right)}\;,
    \label{eq:constraints}
\end{equation}
where $\avg{x^k}$ is the $k$th moment of the sequence $\bm{x}$, and where it is assumed that  the elements of $\bm{d}$ and $\bm{s}$ do not grow with $n$ (i.e., $B$ is sparse).
Thus, based on the presence of multi-edges alone, there is a stringent upper bound on the fraction of bipartite graphs that are actually in the support of the SCM.
An even smaller fraction remains after the bipartite graph with included neighborhood are removed.

\subsection{Markov chain Monte Carlo method}
To sample from the SCM, then, one needs to sample uniformly from a very constrained space, i.e., that of all sequence--preserving bipartite graphs with joint degree sequence $(\bm{d},\bm{s})$.
Previously proposed  approaches such as rejection sampling do not work well \cite{courtney2016generalized}, because natural proposal distributions (e.g., stub matching) give an appreciable weight to non-sequence-preserving bipartite graphs [see Eq.~\eqref{eq:constraints}].
Thus, we turn to the Markov chain Monte Carlo (MCMC) sampling strategy \footnote{We provide a reference \texttt{c++} implementation of the sampler as well as tutorials at \url{https://www.github.com/jg-you/scm}}, which has been used with great success for the CM \cite{fosdick2016configuring,miklos2013towards}.
The general idea is to construct a random chain of sequence--preserving bipartite graphs $B_0,\hdots ,B_T$, to sample from this chain at regular intervals, and to treat the samples as if they had been drawn identically and independently from the ensemble.
The algorithm will be correct if the chain is ergodic (time averages equal ensemble averages)  and uniform  (all non isomorphic $B$ are represented equally).
These properties are determined by the allowed transformations $B_t\to B_{t+1}$ and the resulting transition matrix $\bm{\pi}$, where $\pi_{ij}$ is the probability that $B_j$ follows $B_i$ in the chain.
If the move set \emph{connects the space} and the chain is \emph{aperiodic}, then the chain will be ergodic.
If the transition matrix is \emph{doubly stochastic} (all rows and columns sum to 1), then the chain will be uniform.

We claim that the following set of moves satisfies all three conditions.
Consider $L$, a random variable on the support $\mathcal{L}=\{2,3,\hdots ,L_{\max}\}$, where $L_{\max}$ is a parameter and the distribution $\mathbb{P}[L=\ell]$ is arbitrary but non zero everywhere on $\mathcal{L}$ (for illustration purposes, we will use $\mathbb{P}[L=\ell;\lambda]\propto e^{-\lambda \ell}$).
At each step of the chain, we pick $L$ edges in $B$ (uniformly at random).
We cut these edges and randomly match the stubs stemming from facets to the stubs stemming from nodes.
If this matching generates a sequence--preserving bipartite graph $B'$, then we accept the move; otherwise we resample $B$.
This set of moves is similar to the double-edge swap commonly used in graph MCMC \cite{fosdick2016configuring}.
The only difference is the variable number of rewired edges, added to help the sampler better navigate the constrained support \cite{miklos2013towards}.
Much like its graphical counterpart, the resulting MCMC algorithm is efficient since drawing $L$ edges and checking for resampling can be done in polynomial times.

The chain is aperiodic because the above set of moves yields a doubly stochastic transition matrix for any distribution $\mathbb{P}$: The total number of possible transitions at each configuration is a constant independent from the configuration considered (resampling guarantees this) \cite{fosdick2016configuring}.
The chain is also aperiodic, because there exists orbits of period 1 (resampling steps) and 2 (all moves are reversible) for any nontrivial $(\bm{d},\bm{s})$.

\begin{figure}
    \centering
    \includegraphics[width=\linewidth]{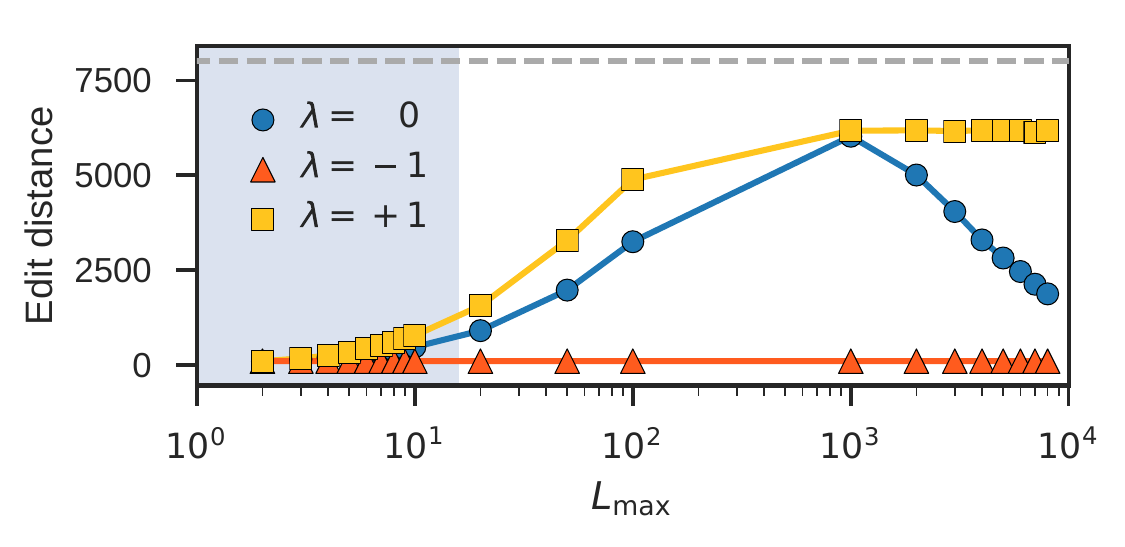}
    \caption{Effect of the parametrization of the proposal distribution $\mathbb{P}$ on the mixing time, as quantified by the edit distances of the graphical representation of the samples.
    We investigate the family of distributions $\mathbb{P}[L=\ell;\lambda]=e^{\lambda \ell}/Z$, and use the regular SCM of $f=1\,000$ facets of size $s=8$, and $n=2\,000$ nodes of degrees $d=4$.
    Pairs of samples are separated by $100$ proposed MCMC moves, and are obtained from a unique initial configuration found via rejection sampling.
    The shaded region lies below the upper bound on $L_{\max}^*$ of Eq.~\eqref{eq:critical_bound}.
    $\lambda=1$ balances high-rejection probability but efficient moves with safe but inefficient moves, yielding the best overall performance for all $L_{\max}$.
    In practice, we have found that medium values of $L_{\max}$ are better, because checking for resampling is of complexity $O(L_{\max}\avg{d})$, which translates into  slower effective mixing time when $L_{\max}\gg 1$.
    }
    \label{fig:change_in_proposal}
\end{figure}

This leaves open the question of whether the support of the SCM is connected by the set of moves or not. 
We argue that it is, for all $L_{\max}\geq L_{\max}^*$, where $L_{\max}^*$ is bounded by
\begin{equation}
    \label{eq:critical_bound}
    L_{\max}^*\leq 2\max\bm{s}\;.
\end{equation}
To prove this, one would have to show that given two sequence--preserving bipartite graphs $B_1$ and $B_2$, it is always possible to find a $B_3$ such that 
$\big|\Delta_{+}\bigl(K(B_1),K(B_2)\bigr)|\geq|\Delta_{+}\bigl(K(B_1),K(B_3)\bigr)\big|$,
where $\Delta_{+}$ is the set of facets in $K(B_2)$ that are not in $K(B_1)$, and $\Delta_{-}$ is the set of facets in $K(B_1)$ that are not in $K(B_2)$ [$K(B)$ is the simplicial complex associated to the graph $B$].
Although a general proof remains elusive, we propose the following non-rigorous argument, valid for sparse simplicial complexes (simplicial complexes with bounded $\max\bm{d}$ and $\max\bm{s}$ in the limit $n\to\infty$).

To construct $B_3$, we first select a facet $\sigma$ in $\Delta_{+}$ (incident on the set of nodes $\Sigma$ in $B_2$).
The conservation of sizes and degrees guarantees that there exists a facet $\tau\in\Delta_{-}$ of the same size.
The idea is then to start from $B_1$, cut all edges attached to $\tau$ and one edge from every node in $\Sigma$, match the stubs of $\sigma$ to those of $v\in\Sigma$, and finally match the remaining orphaned stubs.
This algorithm ensures that $B_3$ is closer to $B_2$ than $B_1$ was, because it removes facets from $\Delta_{\pm}$ (and does not add new facets either: Each $v\in\Sigma$ has at least on facet in $\Delta_{-}$ by the conservation of degrees).
In general, it is not guaranteed that the last step can be carried out without creating included faces.
However, in sparse simplicial complexes, $\sigma$ is well separated from $\tau$ for almost all $(\sigma,\tau)$, since $B_1$ is locally treelike \cite{newman2001random}.
In such cases, no included faces are created at the last step, and the above algorithm can be carried through for some $(\sigma,\tau)$, generating $B_3$.
Because this scheme involves at most $L_{\max}^*=2\max\bm{s}$ rewired edges (when $|\tau|=|\sigma|=\max\bm{s}$), we obtain the  bound  of Eq.~\eqref{eq:critical_bound} for infinite sparse SCM.
In practice, $L_{\max}=2$ seems to always connect the space (we found no counterexamples), and sampling is more efficient when $L_{\max}\gg 2$ (see Fig.~\ref{fig:change_in_proposal})---the value of $L_{\max}^*$ is thus more of theoretical than practical interest.

\begin{figure}
    \includegraphics[width=\linewidth]{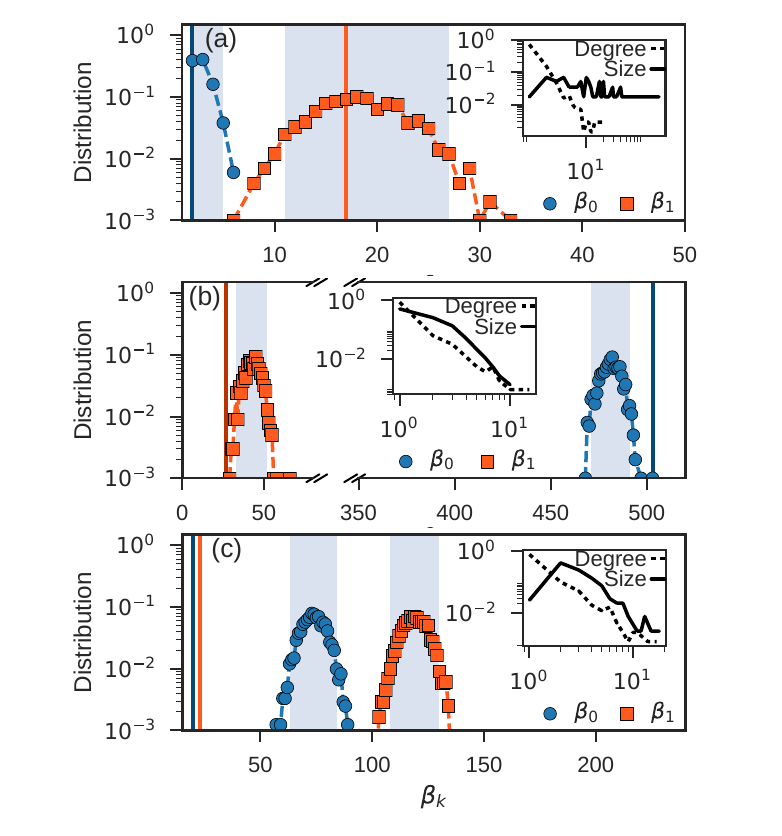}
    \caption{Significance of the Betti numbers of real systems.
    The datasets are bipartite networks, which we  convert to simplicial complexes (we prune included faces).
    They map the relationships between (a) flower-visiting insects (nodes, $n=679$) and plants (facets $f=57$) in Kyoto \cite{kato1990insect}, (b) human disease (nodes $n=1100$) and genes (facets $f=752$) linked by known disorder–gene associations \cite{goh2007human}, and (c) crimes (nodes, $n=829$) and suspects, victims and witnesses (facets, $f=378$) in St. Louis \cite{decker1991}.
    The Betti numbers of these real systems appear as solid vertical lines, and are equal to (a) $\beta_0=2$, $\beta_1=17$ (b) $\beta_0=503,\,\beta_1=27,$ and (c) $\beta_0=20,\,\beta_1=23$.
    We show the distributions of Betti numbers for the equivalent SCM with solid symbols (computed from $1000$ instances of the model).
    The shaded regions contain $95\%$ of the samples.
    The parameters of the SCM---extracted from real systems---are shown in insets.
    }
    \label{fig:case_study}
\end{figure}

\section{Null model}
\label{section:null_model}
We put our efficient MCMC algorithm to the test, by verifying the statistical significance of the structural patterns found in three relational datasets that can be represented as simplicial complexes (see caption of Fig.~\ref{fig:case_study} for details).

Since an instance of the SCM is provided in each case (the real system), we use it as the initial condition for each independent run of the sampling algorithm.
Ergodicity implies that the state of the sampler will be uncorrelated with the initial configuration after a sufficiently long burn-in period---the choice of initial condition is ultimately irrelevant.
Extrapolating from the results of Fig.~\ref{fig:change_in_proposal}, we opt for the proposal distribution $\mathbb{P}[L=\ell]=e^{\lambda\ell}/Z$  with $\lambda=1$ and $L_{\max}$ set to $10\%$ of $m=\sum d_i =\sum s_i$.
Non rigorous arguments from expander graph theory suggest $t_f =O(m\log m)$ as a good---if overzealous---choice of sampling interval \footnote{We represent the support of the SCM as a graph $\mathcal{G}(L_{\max})=(V,E)$. If $\mathcal{G}$ is an expander, then the sampler yields uncorrelated configuration with high probability after $t_f=O(\log |V|)$ steps; the suggested $t_f$ follows from the loose upper bound $|V| \leq m!$. A proof that $\mathcal{G}(L_{\max})$ is in fact an expander will depend on the specifics of $(\bm{d},\bm{s},L_{\max})$; however, we note that $\mathcal{G}(L_{\max})$ shares the two fundamental properties with all expanders for sufficiently large $L_{\max}$:  It is connected and not bipartite.}.

Significance results only make sense if they rely on a null model that embodies a natural null hypothesis for the problem at hand \cite{fosdick2016configuring}.
For example, the regular CM and its correlated variants usefully show that the network projection of datasets with high order interactions are abnormally clustered \cite{newman2003properties}.
Therefore, we use the sampler to investigate the distribution of a mesoscopic property only accessible when the datasets are encoded as simplicial complexes: The \emph{shape} of the datasets, as captured by their homology, i.e., the pattern of holes, cavities and higher dimensional voids \cite{hatcher2000algebraic}.
The homology can be summarized by a series of Betti numbers $\bm{\beta}=(\beta_0,\beta_1,\beta_2,\hdots )$, where $\beta_k$ counts the number of structural holes bounded by $k$-dimensional simplices.
For example, $\beta_0$ counts the number of connected component, $\beta_1$ the number of homological cycles in $K$, $\beta_2$ the number of holes enclosed by facets of sizes $2$, etc.
Since every instance of the SCM has the same fixed local structure but is otherwise maximally random, we expect significant  differences between the Betti number $\bm{\beta}$ of an organized simplicial complex and the bulk of the distribution of $\bm{\beta}$ in the corresponding randomized ensembles.

We show in Fig.~\ref{fig:case_study} the distribution of $\beta_0$ and $\beta_1$ for the SCM associated to the real systems.
Looking first at $\beta_0$, we find that the structure of the pollinator dataset is essentially random [Fig.~\ref{fig:case_study}(a)].
That is, the overwhelming majority of simplicial complexes with the same sequences have similar $\beta_0$.
In contrast, the $\beta_0$ of the disease genome regulation (hereafter \emph{diseasome}) and crime complexes are highly significant [Fig.~\ref{fig:case_study}~(b)--(c)]: A random instance of the SCM has fewer (diseasome) or more (crime) components than the real system with high probability.
In one case (crime), the difference is a statistical signature of how the dataset was gathered, namely by looking up the ties of suspects, victims and witnesses already in the dataset, recursively \cite{decker1991}.
Because this process creates much larger connected components than random sampling, the resulting $\beta_0$ is far from the ensemble average---an effect that we expect to find in any dataset constructed using a similar methodology.
In the other case (diseasome), the real system has \emph{more} components than one would typically expect from the local information alone.
The construction procedure does not explain this disparity \cite{goh2007human}, meaning that the system must self-organize in a fragmented way, likely for biological or evolutionary reasons.

Turning to $\beta_1$ we again find that the structure of the pollinator dataset is typical, and that the same cannot be said of the diseasome and crime datasets.
Both simplicial complexes have significantly fewer cycles than expected; i.e., given a cycle, it is more likely to be filled by a simplex in the real system than in the randomized one, suggesting that some form of high order triadic closure is at play \cite{zuev2015exponential}.
The difference is, however, much more pronounced in the crime dataset; this could be due to the fact that it describes a social system, whose structure tend to be heavily driven by triadic closure  \cite{hidalgo2016disconnected} (and potential high order analogs).

Finally, taking both distributions into account, we conclude that the shape of the pollinator dataset is completely determined by its local structure, while large--scale organizational principles influence the structure of the other datasets.
This leads us to two final observations:
One, care must be exerted in drawing conclusions about the shape of complex datasets---from the homology point of view there is nothing of note in the structure of the pollinator dataset.
Two, some datasets---here the crime and diseasome datasets---are decidedly \emph{not} random.
This raises the question of just how much information must models account for, before they can capture such atypical Betti numbers.
Would, for example, adding limited correlations among degrees be sufficient to capture the shape of most real datasets?
Or do we need to embrace growth models, with their sophisticated rules, and clustered local structure \cite{wu2015emergent,bianconi2016network,hebert2015complex}?\\

\section{Perspectives}
As it stands, the SCM already establishes the analysis of simplicial complexes on firmer statistical ground.
The next step will be to clarify a number of important open questions, e.g., what is the true value of $L_{\max}^*$  for arbitrary simplicial complexes, and what is optimal choice of proposal distribution $\mathbb{P}$ (cf. Fig.~\ref{fig:change_in_proposal}).

Beyond these obvious questions, the connection between the SCM and the simple CM lead us to a series of natural problems not addressed in this paper.
These include the problem of the \emph{simpliciality} of arbitrary pairs of sequences (i.e., is there a simplicial complex which realize a pair of sequences?) \cite{courtney2016generalized}, related to the problem of constructing initial conditions for the MCMC sampler, when no real system is available.
We believe that the solution to such problems will require new insights, as the no--inclusion constraint appears to be a major obstacle to the application of classical methods developed for the analogous \emph{graphicality} problem \cite{havel1955remark,hakimi1962realizability}.

In closing, we stress that all the above questions and challenges are of technical nature; the model and sampler can already be applied to practical problems \cite{Note1}.
This could lead to improvements in persistent homology (e.g. statistically sound filtrations of weighted complexes) or a formulation of community detection of simplicial complexes (via modularity \cite{newman2004finding}), and could provide a new glimpse into the emergence of homology and higher order structural properties in real complex systems.

\section*{Acknowledgments}
We thank L. H\'ebert-Dufresne  and G. Bianconi for helpful discussions and comments.
The authors acknowledge the support of the ADnD project by Compagnia San Paolo (AP, GP), the Fonds de recherche du Qu\'ebec-Nature et technologies (JGY), the Complex Systems Langrange Lab (FV), and the YRNCS Bridge Grant (AP, JGY).
AP is grateful for the hospitality of L.J.~Dub\'e at Universit\'e Laval, where parts of the research work was conducted.\\
AP and JGY contributed equally to this work.


\begin{thebibliography}{41}%
\makeatletter
\providecommand \@ifxundefined [1]{%
 \@ifx{#1\undefined}
}%
\providecommand \@ifnum [1]{%
 \ifnum #1\expandafter \@firstoftwo
 \else \expandafter \@secondoftwo
 \fi
}%
\providecommand \@ifx [1]{%
 \ifx #1\expandafter \@firstoftwo
 \else \expandafter \@secondoftwo
 \fi
}%
\providecommand \natexlab [1]{#1}%
\providecommand \enquote  [1]{``#1''}%
\providecommand \bibnamefont  [1]{#1}%
\providecommand \bibfnamefont [1]{#1}%
\providecommand \citenamefont [1]{#1}%
\providecommand \href@noop [0]{\@secondoftwo}%
\providecommand \href [0]{\begingroup \@sanitize@url \@href}%
\providecommand \@href[1]{\@@startlink{#1}\@@href}%
\providecommand \@@href[1]{\endgroup#1\@@endlink}%
\providecommand \@sanitize@url [0]{\catcode `\\12\catcode `\$12\catcode
  `\&12\catcode `\#12\catcode `\^12\catcode `\_12\catcode `\%12\relax}%
\providecommand \@@startlink[1]{}%
\providecommand \@@endlink[0]{}%
\providecommand \url  [0]{\begingroup\@sanitize@url \@url }%
\providecommand \@url [1]{\endgroup\@href {#1}{\urlprefix }}%
\providecommand \urlprefix  [0]{URL }%
\providecommand \Eprint [0]{\href }%
\providecommand \doibase [0]{http://dx.doi.org/}%
\providecommand \selectlanguage [0]{\@gobble}%
\providecommand \bibinfo  [0]{\@secondoftwo}%
\providecommand \bibfield  [0]{\@secondoftwo}%
\providecommand \translation [1]{[#1]}%
\providecommand \BibitemOpen [0]{}%
\providecommand \bibitemStop [0]{}%
\providecommand \bibitemNoStop [0]{.\EOS\space}%
\providecommand \EOS [0]{\spacefactor3000\relax}%
\providecommand \BibitemShut  [1]{\csname bibitem#1\endcsname}%
\let\auto@bib@innerbib\@empty
\bibitem [{\citenamefont {Pastor-Satorras}\ \emph {et~al.}(2015)\citenamefont
  {Pastor-Satorras}, \citenamefont {Castellano}, \citenamefont {Van~Mieghem},\
  and\ \citenamefont {Vespignani}}]{pastor2015epidemic}%
  \BibitemOpen
  \bibfield  {author} {\bibinfo {author} {\bibfnamefont {R.}~\bibnamefont
  {Pastor-Satorras}}, \bibinfo {author} {\bibfnamefont {C.}~\bibnamefont
  {Castellano}}, \bibinfo {author} {\bibfnamefont {P.}~\bibnamefont
  {Van~Mieghem}}, \ and\ \bibinfo {author} {\bibfnamefont {A.}~\bibnamefont
  {Vespignani}},\ }\href {\doibase 10.1103/RevModPhys.87.925} {\bibfield
  {journal} {\bibinfo  {journal} {Rev. Mod. Phys.}\ }\textbf {\bibinfo {volume}
  {87}},\ \bibinfo {pages} {925} (\bibinfo {year} {2015})}\BibitemShut
  {NoStop}%
\bibitem [{\citenamefont {Liu}\ and\ \citenamefont
  {Barab{\'a}si}(2016)}]{liu2016control}%
  \BibitemOpen
  \bibfield  {author} {\bibinfo {author} {\bibfnamefont {Y.-Y.}\ \bibnamefont
  {Liu}}\ and\ \bibinfo {author} {\bibfnamefont {A.-L.}\ \bibnamefont
  {Barab{\'a}si}},\ }\href {\doibase 10.1103/RevModPhys.88.035006} {\bibfield
  {journal} {\bibinfo  {journal} {Rev. Mod. Phys.}\ }\textbf {\bibinfo {volume}
  {88}},\ \bibinfo {pages} {035006} (\bibinfo {year} {2016})}\BibitemShut
  {NoStop}%
\bibitem [{\citenamefont {Porter}\ \emph {et~al.}(2009)\citenamefont {Porter},
  \citenamefont {Onnela},\ and\ \citenamefont {Mucha}}]{porter2009communities}%
  \BibitemOpen
  \bibfield  {author} {\bibinfo {author} {\bibfnamefont {M.~A.}\ \bibnamefont
  {Porter}}, \bibinfo {author} {\bibfnamefont {J.-P.}\ \bibnamefont {Onnela}},
  \ and\ \bibinfo {author} {\bibfnamefont {P.~J.}\ \bibnamefont {Mucha}},\
  }\href {http://people.maths.ox.ac.uk/~porterm/papers/comnotices.pdf}
  {\bibfield  {journal} {\bibinfo  {journal} {Notices AMS}\ }\textbf {\bibinfo
  {volume} {56}},\ \bibinfo {pages} {1082} (\bibinfo {year}
  {2009})}\BibitemShut {NoStop}%
\bibitem [{\citenamefont {Newman}(2012)}]{newman2012communities}%
  \BibitemOpen
  \bibfield  {author} {\bibinfo {author} {\bibfnamefont {M.~E.~J.}\
  \bibnamefont {Newman}},\ }\href {\doibase 10.1038/nphys2162} {\bibfield
  {journal} {\bibinfo  {journal} {Nat. Phys.}\ }\textbf {\bibinfo {volume}
  {8}},\ \bibinfo {pages} {25} (\bibinfo {year} {2012})}\BibitemShut {NoStop}%
\bibitem [{\citenamefont {Dabaghian}\ \emph {et~al.}(2012)\citenamefont
  {Dabaghian}, \citenamefont {M{\'e}moli}, \citenamefont {Frank},\ and\
  \citenamefont {Carlsson}}]{dabaghian2012topological}%
  \BibitemOpen
  \bibfield  {author} {\bibinfo {author} {\bibfnamefont {Y.}~\bibnamefont
  {Dabaghian}}, \bibinfo {author} {\bibfnamefont {F.}~\bibnamefont
  {M{\'e}moli}}, \bibinfo {author} {\bibfnamefont {L.}~\bibnamefont {Frank}}, \
  and\ \bibinfo {author} {\bibfnamefont {G.}~\bibnamefont {Carlsson}},\ }\href
  {\doibase 10.1371/journal.pcbi.1002581} {\bibfield  {journal} {\bibinfo
  {journal} {PLoS Comput. Biol.}\ }\textbf {\bibinfo {volume} {8}},\ \bibinfo
  {pages} {e1002581} (\bibinfo {year} {2012})}\BibitemShut {NoStop}%
\bibitem [{\citenamefont {Giusti}\ \emph {et~al.}(2016)\citenamefont {Giusti},
  \citenamefont {Ghrist},\ and\ \citenamefont {Bassett}}]{giusti2016two}%
  \BibitemOpen
  \bibfield  {author} {\bibinfo {author} {\bibfnamefont {C.}~\bibnamefont
  {Giusti}}, \bibinfo {author} {\bibfnamefont {R.}~\bibnamefont {Ghrist}}, \
  and\ \bibinfo {author} {\bibfnamefont {D.~S.}\ \bibnamefont {Bassett}},\
  }\href {\doibase 10.1007/s10827-016-0608-6} {\bibfield  {journal} {\bibinfo
  {journal} {J. Comput. Neurosci.}\ }\textbf {\bibinfo {volume} {41}},\
  \bibinfo {pages} {1} (\bibinfo {year} {2016})}\BibitemShut {NoStop}%
\bibitem [{\citenamefont {Xia}\ and\ \citenamefont
  {Wei}(2014)}]{xia2014persistent}%
  \BibitemOpen
  \bibfield  {author} {\bibinfo {author} {\bibfnamefont {K.}~\bibnamefont
  {Xia}}\ and\ \bibinfo {author} {\bibfnamefont {G.-W.}\ \bibnamefont {Wei}},\
  }\href {\doibase 10.1002/cnm.2655} {\bibfield  {journal} {\bibinfo  {journal}
  {Int. J. Numer. Methods Biomed. Eng.}\ }\textbf {\bibinfo {volume} {30}},\
  \bibinfo {pages} {814} (\bibinfo {year} {2014})}\BibitemShut {NoStop}%
\bibitem [{\citenamefont {H{\'e}bert-Dufresne}\ \emph
  {et~al.}(2015)\citenamefont {H{\'e}bert-Dufresne}, \citenamefont {Laurence},
  \citenamefont {Allard}, \citenamefont {Young},\ and\ \citenamefont
  {Dub{\'e}}}]{hebert2015complex}%
  \BibitemOpen
  \bibfield  {author} {\bibinfo {author} {\bibfnamefont {L.}~\bibnamefont
  {H{\'e}bert-Dufresne}}, \bibinfo {author} {\bibfnamefont {E.}~\bibnamefont
  {Laurence}}, \bibinfo {author} {\bibfnamefont {A.}~\bibnamefont {Allard}},
  \bibinfo {author} {\bibfnamefont {J.-G.}\ \bibnamefont {Young}}, \ and\
  \bibinfo {author} {\bibfnamefont {L.~J.}\ \bibnamefont {Dub{\'e}}},\ }\href
  {\doibase 10.1103/PhysRevE.92.062809} {\bibfield  {journal} {\bibinfo
  {journal} {Phys. Rev. E}\ }\textbf {\bibinfo {volume} {92}},\ \bibinfo
  {pages} {062809} (\bibinfo {year} {2015})}\BibitemShut {NoStop}%
\bibitem [{\citenamefont {Stolz}\ \emph {et~al.}(shed)\citenamefont {Stolz},
  \citenamefont {Harrington},\ and\ \citenamefont
  {Porter}}]{stolz2016topological}%
  \BibitemOpen
  \bibfield  {author} {\bibinfo {author} {\bibfnamefont {B.}~\bibnamefont
  {Stolz}}, \bibinfo {author} {\bibfnamefont {H.}~\bibnamefont {Harrington}}, \
  and\ \bibinfo {author} {\bibfnamefont {M.~A.}\ \bibnamefont {Porter}},\
  }\href {http://arxiv.org/abs/1610.00752} {\bibfield  {journal} {\bibinfo
  {journal} {arXiv:1610.00752}\ } (\bibinfo {year} {unpublished})}\BibitemShut
  {NoStop}%
\bibitem [{\citenamefont {Zuev}\ \emph {et~al.}(2015)\citenamefont {Zuev},
  \citenamefont {Eisenberg},\ and\ \citenamefont
  {Krioukov}}]{zuev2015exponential}%
  \BibitemOpen
  \bibfield  {author} {\bibinfo {author} {\bibfnamefont {K.}~\bibnamefont
  {Zuev}}, \bibinfo {author} {\bibfnamefont {O.}~\bibnamefont {Eisenberg}}, \
  and\ \bibinfo {author} {\bibfnamefont {D.}~\bibnamefont {Krioukov}},\ }\href
  {\doibase 10.1088/1751-8113/48/46/465002} {\bibfield  {journal} {\bibinfo
  {journal} {J. Phys. A}\ }\textbf {\bibinfo {volume} {48}},\ \bibinfo {pages}
  {465002} (\bibinfo {year} {2015})}\BibitemShut {NoStop}%
\bibitem [{\citenamefont {Ellis}\ and\ \citenamefont
  {Klein}(2014)}]{klein2014}%
  \BibitemOpen
  \bibfield  {author} {\bibinfo {author} {\bibfnamefont {S.~P.}\ \bibnamefont
  {Ellis}}\ and\ \bibinfo {author} {\bibfnamefont {A.}~\bibnamefont {Klein}},\
  }\href {\doibase 10.4310/HHA.2014.v16.n1.a14} {\bibfield  {journal} {\bibinfo
   {journal} {Homology, Homotopy Appl.}\ }\textbf {\bibinfo {volume} {16}},\
  \bibinfo {pages} {245} (\bibinfo {year} {2014})}\BibitemShut {NoStop}%
\bibitem [{\citenamefont {Curto}\ and\ \citenamefont
  {Itskov}(2008)}]{curto2008cell}%
  \BibitemOpen
  \bibfield  {author} {\bibinfo {author} {\bibfnamefont {C.}~\bibnamefont
  {Curto}}\ and\ \bibinfo {author} {\bibfnamefont {V.}~\bibnamefont {Itskov}},\
  }\href {\doibase 10.1371/journal.pcbi.1000205} {\bibfield  {journal}
  {\bibinfo  {journal} {PLoS Comput. Biol.}\ }\textbf {\bibinfo {volume} {4}},\
  \bibinfo {pages} {e1000205} (\bibinfo {year} {2008})}\BibitemShut {NoStop}%
\bibitem [{\citenamefont {Horak}\ \emph {et~al.}(2009)\citenamefont {Horak},
  \citenamefont {Maleti{\'c}},\ and\ \citenamefont
  {Rajkovi{\'c}}}]{horak2009persistent}%
  \BibitemOpen
  \bibfield  {author} {\bibinfo {author} {\bibfnamefont {D.}~\bibnamefont
  {Horak}}, \bibinfo {author} {\bibfnamefont {S.}~\bibnamefont {Maleti{\'c}}},
  \ and\ \bibinfo {author} {\bibfnamefont {M.}~\bibnamefont {Rajkovi{\'c}}},\
  }\href {\doibase 10.1088/1742-5468/2009/03/P03034} {\bibfield  {journal}
  {\bibinfo  {journal} {J. Stat. Mech. Theor. Exp.}\ }\textbf {\bibinfo
  {volume} {2009}},\ \bibinfo {pages} {P03034} (\bibinfo {year}
  {2009})}\BibitemShut {NoStop}%
\bibitem [{\citenamefont {Patania}\ \emph {et~al.}(2017)\citenamefont
  {Patania}, \citenamefont {Vaccarino},\ and\ \citenamefont
  {Petri}}]{patania2017topological}%
  \BibitemOpen
  \bibfield  {author} {\bibinfo {author} {\bibfnamefont {A.}~\bibnamefont
  {Patania}}, \bibinfo {author} {\bibfnamefont {F.}~\bibnamefont {Vaccarino}},
  \ and\ \bibinfo {author} {\bibfnamefont {G.}~\bibnamefont {Petri}},\ }\href
  {\doibase 10.1140/epjds/s13688-017-0104-x} {\bibfield  {journal} {\bibinfo
  {journal} {EPJ Data Sci.}\ }\textbf {\bibinfo {volume} {6}},\ \bibinfo
  {pages} {7} (\bibinfo {year} {2017})}\BibitemShut {NoStop}%
\bibitem [{\citenamefont {Petri}\ \emph {et~al.}(2013)\citenamefont {Petri},
  \citenamefont {Scolamiero}, \citenamefont {Donato},\ and\ \citenamefont
  {Vaccarino}}]{petri2013topological}%
  \BibitemOpen
  \bibfield  {author} {\bibinfo {author} {\bibfnamefont {G.}~\bibnamefont
  {Petri}}, \bibinfo {author} {\bibfnamefont {M.}~\bibnamefont {Scolamiero}},
  \bibinfo {author} {\bibfnamefont {I.}~\bibnamefont {Donato}}, \ and\ \bibinfo
  {author} {\bibfnamefont {F.}~\bibnamefont {Vaccarino}},\ }\href {\doibase
  10.1371/journal.pone.0066506} {\bibfield  {journal} {\bibinfo  {journal}
  {PloS One}\ }\textbf {\bibinfo {volume} {8}},\ \bibinfo {pages} {e66506}
  (\bibinfo {year} {2013})}\BibitemShut {NoStop}%
\bibitem [{\citenamefont {Sizemore}\ \emph {et~al.}(2017)\citenamefont
  {Sizemore}, \citenamefont {Giusti},\ and\ \citenamefont
  {Bassett}}]{sizemore2016classification}%
  \BibitemOpen
  \bibfield  {author} {\bibinfo {author} {\bibfnamefont {A.}~\bibnamefont
  {Sizemore}}, \bibinfo {author} {\bibfnamefont {C.}~\bibnamefont {Giusti}}, \
  and\ \bibinfo {author} {\bibfnamefont {D.~S.}\ \bibnamefont {Bassett}},\
  }\href {\doibase 10.1093/comnet/cnw013} {\bibfield  {journal} {\bibinfo
  {journal} {J. Complex Netw.}\ }\textbf {\bibinfo {volume} {5}},\ \bibinfo
  {pages} {245} (\bibinfo {year} {2017})}\BibitemShut {NoStop}%
\bibitem [{\citenamefont {Chan}\ \emph {et~al.}(2013)\citenamefont {Chan},
  \citenamefont {Carlsson},\ and\ \citenamefont {Rabadan}}]{chan2013topology}%
  \BibitemOpen
  \bibfield  {author} {\bibinfo {author} {\bibfnamefont {J.~M.}\ \bibnamefont
  {Chan}}, \bibinfo {author} {\bibfnamefont {G.}~\bibnamefont {Carlsson}}, \
  and\ \bibinfo {author} {\bibfnamefont {R.}~\bibnamefont {Rabadan}},\ }\href
  {\doibase 10.1073/pnas.1313480110} {\bibfield  {journal} {\bibinfo  {journal}
  {PNAS}\ }\textbf {\bibinfo {volume} {110}},\ \bibinfo {pages} {18566}
  (\bibinfo {year} {2013})}\BibitemShut {NoStop}%
\bibitem [{\citenamefont {Petri}\ \emph {et~al.}(2014)\citenamefont {Petri},
  \citenamefont {Expert}, \citenamefont {Turkheimer}, \citenamefont
  {Carhart-Harris}, \citenamefont {Nutt}, \citenamefont {Hellyer},\ and\
  \citenamefont {Vaccarino}}]{petri2014homological}%
  \BibitemOpen
  \bibfield  {author} {\bibinfo {author} {\bibfnamefont {G.}~\bibnamefont
  {Petri}}, \bibinfo {author} {\bibfnamefont {P.}~\bibnamefont {Expert}},
  \bibinfo {author} {\bibfnamefont {F.}~\bibnamefont {Turkheimer}}, \bibinfo
  {author} {\bibfnamefont {R.}~\bibnamefont {Carhart-Harris}}, \bibinfo
  {author} {\bibfnamefont {D.}~\bibnamefont {Nutt}}, \bibinfo {author}
  {\bibfnamefont {P.}~\bibnamefont {Hellyer}}, \ and\ \bibinfo {author}
  {\bibfnamefont {F.}~\bibnamefont {Vaccarino}},\ }\href {\doibase
  10.1098/rsif.2014.0873} {\bibfield  {journal} {\bibinfo  {journal} {J. R.
  Soc. Interface}\ }\textbf {\bibinfo {volume} {11}},\ \bibinfo {pages}
  {20140873} (\bibinfo {year} {2014})}\BibitemShut {NoStop}%
\bibitem [{\citenamefont {Hiraoka}\ \emph {et~al.}(2016)\citenamefont
  {Hiraoka}, \citenamefont {Nakamura}, \citenamefont {Hirata}, \citenamefont
  {Escolar}, \citenamefont {Matsue},\ and\ \citenamefont
  {Nishiura}}]{hiraoka2016hierarchical}%
  \BibitemOpen
  \bibfield  {author} {\bibinfo {author} {\bibfnamefont {Y.}~\bibnamefont
  {Hiraoka}}, \bibinfo {author} {\bibfnamefont {T.}~\bibnamefont {Nakamura}},
  \bibinfo {author} {\bibfnamefont {A.}~\bibnamefont {Hirata}}, \bibinfo
  {author} {\bibfnamefont {E.~G.}\ \bibnamefont {Escolar}}, \bibinfo {author}
  {\bibfnamefont {K.}~\bibnamefont {Matsue}}, \ and\ \bibinfo {author}
  {\bibfnamefont {Y.}~\bibnamefont {Nishiura}},\ }\href {\doibase
  10.1073/pnas.1520877113} {\bibfield  {journal} {\bibinfo  {journal} {PNAS}\
  }\textbf {\bibinfo {volume} {113}},\ \bibinfo {pages} {7035} (\bibinfo {year}
  {2016})}\BibitemShut {NoStop}%
\bibitem [{\citenamefont {Kahle}(2014)}]{kahle2014topology}%
  \BibitemOpen
  \bibfield  {author} {\bibinfo {author} {\bibfnamefont {M.}~\bibnamefont
  {Kahle}},\ }\href {\doibase 10.1090/conm/620/12367} {\bibfield  {journal}
  {\bibinfo  {journal} {AMS Contemp. Math.}\ }\textbf {\bibinfo {volume}
  {620}},\ \bibinfo {pages} {201} (\bibinfo {year} {2014})}\BibitemShut
  {NoStop}%
\bibitem [{\citenamefont {Costa}\ and\ \citenamefont
  {Farber}(2016)}]{costa2016random}%
  \BibitemOpen
  \bibfield  {author} {\bibinfo {author} {\bibfnamefont {A.}~\bibnamefont
  {Costa}}\ and\ \bibinfo {author} {\bibfnamefont {M.}~\bibnamefont {Farber}},\
  }in\ \href@noop {} {\emph {\bibinfo {booktitle} {Configuration Spaces}}}\
  (\bibinfo  {publisher} {Springer, Berlin},\ \bibinfo {year} {2016})\ pp.\
  \bibinfo {pages} {129--153}\BibitemShut {NoStop}%
\bibitem [{\citenamefont {Courtney}\ and\ \citenamefont
  {Bianconi}(2016)}]{courtney2016generalized}%
  \BibitemOpen
  \bibfield  {author} {\bibinfo {author} {\bibfnamefont {O.~T.}\ \bibnamefont
  {Courtney}}\ and\ \bibinfo {author} {\bibfnamefont {G.}~\bibnamefont
  {Bianconi}},\ }\href {\doibase 10.1103/PhysRevE.93.062311} {\bibfield
  {journal} {\bibinfo  {journal} {Phys. Rev. E}\ }\textbf {\bibinfo {volume}
  {93}},\ \bibinfo {pages} {062311} (\bibinfo {year} {2016})}\BibitemShut
  {NoStop}%
\bibitem [{\citenamefont {Bianconi}\ and\ \citenamefont
  {Rahmede}(2016)}]{bianconi2016network}%
  \BibitemOpen
  \bibfield  {author} {\bibinfo {author} {\bibfnamefont {G.}~\bibnamefont
  {Bianconi}}\ and\ \bibinfo {author} {\bibfnamefont {C.}~\bibnamefont
  {Rahmede}},\ }\href {\doibase 10.1103/PhysRevE.93.032315} {\bibfield
  {journal} {\bibinfo  {journal} {Phys. Rev. E}\ }\textbf {\bibinfo {volume}
  {93}},\ \bibinfo {pages} {032315} (\bibinfo {year} {2016})}\BibitemShut
  {NoStop}%
\bibitem [{\citenamefont {Wu}\ \emph {et~al.}(2015)\citenamefont {Wu},
  \citenamefont {Menichetti}, \citenamefont {Rahmede},\ and\ \citenamefont
  {Bianconi}}]{wu2015emergent}%
  \BibitemOpen
  \bibfield  {author} {\bibinfo {author} {\bibfnamefont {Z.}~\bibnamefont
  {Wu}}, \bibinfo {author} {\bibfnamefont {G.}~\bibnamefont {Menichetti}},
  \bibinfo {author} {\bibfnamefont {C.}~\bibnamefont {Rahmede}}, \ and\
  \bibinfo {author} {\bibfnamefont {G.}~\bibnamefont {Bianconi}},\ }\href
  {\doibase 10.1038/srep10073} {\bibfield  {journal} {\bibinfo  {journal} {Sci.
  Rep.}\ }\textbf {\bibinfo {volume} {5}},\ \bibinfo {pages} {10073} (\bibinfo
  {year} {2015})}\BibitemShut {NoStop}%
\bibitem [{\citenamefont {Fosdick}\ \emph {et~al.}(shed)\citenamefont
  {Fosdick}, \citenamefont {Larremore}, \citenamefont {Nishimura},\ and\
  \citenamefont {Ugander}}]{fosdick2016configuring}%
  \BibitemOpen
  \bibfield  {author} {\bibinfo {author} {\bibfnamefont {B.~K.}\ \bibnamefont
  {Fosdick}}, \bibinfo {author} {\bibfnamefont {D.~B.}\ \bibnamefont
  {Larremore}}, \bibinfo {author} {\bibfnamefont {J.}~\bibnamefont
  {Nishimura}}, \ and\ \bibinfo {author} {\bibfnamefont {J.}~\bibnamefont
  {Ugander}},\ }\href {https://arxiv.org/abs/1608.00607} {\bibfield  {journal}
  {\bibinfo  {journal} {arXiv:1608.00607}\ } (\bibinfo {year}
  {unpublished})}\BibitemShut {NoStop}%
\bibitem [{\citenamefont {Orsini}\ \emph {et~al.}(2015)\citenamefont {Orsini},
  \citenamefont {Dankulov}, \citenamefont {Jamakovic}, \citenamefont
  {Mahadevan}, \citenamefont {Colomer-de Sim{\'o}n}, \citenamefont {Vahdat},
  \citenamefont {Bassler}, \citenamefont {Toroczkai}, \citenamefont
  {Bogu{\~n}{\'a}}, \citenamefont {Caldarelli} \emph
  {et~al.}}]{orsini2015quantifying}%
  \BibitemOpen
  \bibfield  {author} {\bibinfo {author} {\bibfnamefont {C.}~\bibnamefont
  {Orsini}}, \bibinfo {author} {\bibfnamefont {M.~M.}\ \bibnamefont
  {Dankulov}}, \bibinfo {author} {\bibfnamefont {A.}~\bibnamefont {Jamakovic}},
  \bibinfo {author} {\bibfnamefont {P.}~\bibnamefont {Mahadevan}}, \bibinfo
  {author} {\bibfnamefont {P.}~\bibnamefont {Colomer-de Sim{\'o}n}}, \bibinfo
  {author} {\bibfnamefont {A.}~\bibnamefont {Vahdat}}, \bibinfo {author}
  {\bibfnamefont {K.~E.}\ \bibnamefont {Bassler}}, \bibinfo {author}
  {\bibfnamefont {Z.}~\bibnamefont {Toroczkai}}, \bibinfo {author}
  {\bibfnamefont {M.}~\bibnamefont {Bogu{\~n}{\'a}}}, \bibinfo {author}
  {\bibfnamefont {G.}~\bibnamefont {Caldarelli}},  \emph {et~al.},\ }\href
  {\doibase 10.1038/ncomms9627} {\bibfield  {journal} {\bibinfo  {journal}
  {Nat. Commun.}\ }\textbf {\bibinfo {volume} {6}},\ \bibinfo {pages} {8627}
  (\bibinfo {year} {2015})}\BibitemShut {NoStop}%
\bibitem [{\citenamefont {Molloy}\ and\ \citenamefont
  {Reed}(1995)}]{molloy1995critical}%
  \BibitemOpen
  \bibfield  {author} {\bibinfo {author} {\bibfnamefont {M.}~\bibnamefont
  {Molloy}}\ and\ \bibinfo {author} {\bibfnamefont {B.~A.}\ \bibnamefont
  {Reed}},\ }\href {\doibase 10.1002/rsa.3240060204} {\bibfield  {journal}
  {\bibinfo  {journal} {Rand. Struct. Alg.}\ }\textbf {\bibinfo {volume} {6}},\
  \bibinfo {pages} {161} (\bibinfo {year} {1995})}\BibitemShut {NoStop}%
\bibitem [{\citenamefont {Newman}\ \emph {et~al.}(2001)\citenamefont {Newman},
  \citenamefont {Strogatz},\ and\ \citenamefont {Watts}}]{newman2001random}%
  \BibitemOpen
  \bibfield  {author} {\bibinfo {author} {\bibfnamefont {M.~E.~J.}\
  \bibnamefont {Newman}}, \bibinfo {author} {\bibfnamefont {S.~H.}\
  \bibnamefont {Strogatz}}, \ and\ \bibinfo {author} {\bibfnamefont {D.~J.}\
  \bibnamefont {Watts}},\ }\href {\doibase 10.1103/PhysRevE.64.026118}
  {\bibfield  {journal} {\bibinfo  {journal} {Phys. Rev. E}\ }\textbf {\bibinfo
  {volume} {64}},\ \bibinfo {pages} {026118} (\bibinfo {year}
  {2001})}\BibitemShut {NoStop}%
\bibitem [{\citenamefont {Hatcher}(2000)}]{hatcher2000algebraic}%
  \BibitemOpen
  \bibfield  {author} {\bibinfo {author} {\bibfnamefont {A.}~\bibnamefont
  {Hatcher}},\ }\href@noop {} {\emph {\bibinfo {title} {Algebraic
  {Topology}}}}\ (\bibinfo  {publisher} {Cambridge University Press},\ \bibinfo
  {address} {Cambridge, UK},\ \bibinfo {year} {2000})\BibitemShut {NoStop}%
\bibitem [{\citenamefont {Mikl{\'o}s}\ \emph {et~al.}(2013)\citenamefont
  {Mikl{\'o}s}, \citenamefont {Erd{\H{o}}s},\ and\ \citenamefont
  {Soukup}}]{miklos2013towards}%
  \BibitemOpen
  \bibfield  {author} {\bibinfo {author} {\bibfnamefont {I.}~\bibnamefont
  {Mikl{\'o}s}}, \bibinfo {author} {\bibfnamefont {P.~L.}\ \bibnamefont
  {Erd{\H{o}}s}}, \ and\ \bibinfo {author} {\bibfnamefont {L.}~\bibnamefont
  {Soukup}},\ }\href
  {http://www.combinatorics.org/ojs/index.php/eljc/article/view/3028}
  {\bibfield  {journal} {\bibinfo  {journal} {Electron. J. Combin.}\ }\textbf
  {\bibinfo {volume} {20}},\ \bibinfo {pages} {P16} (\bibinfo {year}
  {2013})}\BibitemShut {NoStop}%
\bibitem [{\citenamefont {Bender}\ and\ \citenamefont
  {Canfield}(1978)}]{bender1978asymptotic}%
  \BibitemOpen
  \bibfield  {author} {\bibinfo {author} {\bibfnamefont {E.~A.}\ \bibnamefont
  {Bender}}\ and\ \bibinfo {author} {\bibfnamefont {E.~R.}\ \bibnamefont
  {Canfield}},\ }\href {\doibase 10.1016/0097-3165(78)90059-6} {\bibfield
  {journal} {\bibinfo  {journal} {J. Combin. Theo. A}\ }\textbf {\bibinfo
  {volume} {24}},\ \bibinfo {pages} {296} (\bibinfo {year} {1978})}\BibitemShut
  {NoStop}%
\bibitem [{Note1()}]{Note1}%
  \BibitemOpen
  \bibinfo {note} {We provide a reference \protect \texttt {c++} implementation
  of the sampler as well as tutorials at \protect \url
  {https://www.github.com/jg-you/scm}}\BibitemShut {NoStop}%
\bibitem [{\citenamefont {Kato}\ \emph {et~al.}(1990)\citenamefont {Kato},
  \citenamefont {Kakutani}, \citenamefont {Inoue},\ and\ \citenamefont
  {Itino}}]{kato1990insect}%
  \BibitemOpen
  \bibfield  {author} {\bibinfo {author} {\bibfnamefont {M.}~\bibnamefont
  {Kato}}, \bibinfo {author} {\bibfnamefont {T.}~\bibnamefont {Kakutani}},
  \bibinfo {author} {\bibfnamefont {T.}~\bibnamefont {Inoue}}, \ and\ \bibinfo
  {author} {\bibfnamefont {T.}~\bibnamefont {Itino}},\ }\href
  {https://repository.kulib.kyoto-u.ac.jp/dspace/bitstream/2433/156101/1/cbl02704_309.pdf}
  {\bibfield  {journal} {\bibinfo  {journal} {Contr. Biol. Lab. Kyoto Univ.}\
  }\textbf {\bibinfo {volume} {27}},\ \bibinfo {pages} {309} (\bibinfo {year}
  {1990})}\BibitemShut {NoStop}%
\bibitem [{\citenamefont {Goh}\ \emph {et~al.}(2007)\citenamefont {Goh},
  \citenamefont {Cusick}, \citenamefont {Valle}, \citenamefont {Childs},
  \citenamefont {Vidal},\ and\ \citenamefont {Barab{\'a}si}}]{goh2007human}%
  \BibitemOpen
  \bibfield  {author} {\bibinfo {author} {\bibfnamefont {K.-I.}\ \bibnamefont
  {Goh}}, \bibinfo {author} {\bibfnamefont {M.~E.}\ \bibnamefont {Cusick}},
  \bibinfo {author} {\bibfnamefont {D.}~\bibnamefont {Valle}}, \bibinfo
  {author} {\bibfnamefont {B.}~\bibnamefont {Childs}}, \bibinfo {author}
  {\bibfnamefont {M.}~\bibnamefont {Vidal}}, \ and\ \bibinfo {author}
  {\bibfnamefont {A.-L.}\ \bibnamefont {Barab{\'a}si}},\ }\href {\doibase
  10.1073/pnas.0701361104} {\bibfield  {journal} {\bibinfo  {journal} {PNAS}\
  }\textbf {\bibinfo {volume} {104}},\ \bibinfo {pages} {8685} (\bibinfo {year}
  {2007})}\BibitemShut {NoStop}%
\bibitem [{\citenamefont {Decker}\ \emph {et~al.}(1991)\citenamefont {Decker},
  \citenamefont {Kohfeld}, \citenamefont {Rosenfeld},\ and\ \citenamefont
  {Sprague}}]{decker1991}%
  \BibitemOpen
  \bibfield  {author} {\bibinfo {author} {\bibfnamefont {S.}~\bibnamefont
  {Decker}}, \bibinfo {author} {\bibfnamefont {C.~W.}\ \bibnamefont {Kohfeld}},
  \bibinfo {author} {\bibfnamefont {R.}~\bibnamefont {Rosenfeld}}, \ and\
  \bibinfo {author} {\bibfnamefont {J.}~\bibnamefont {Sprague}},\ }\href@noop
  {} {\emph {\bibinfo {title} {St. {Louis} Homicide Project: Local Responses to
  a National Problem}}}\ (\bibinfo  {publisher} {University of Missouri},\
  \bibinfo {address} {St. Louis},\ \bibinfo {year} {1991})\BibitemShut
  {NoStop}%
\bibitem [{Note2()}]{Note2}%
  \BibitemOpen
  \bibinfo {note} {We represent the support of the SCM as a graph $\protect
  \mathcal {G}(L_{\protect \qopname \relax m{max}})=(V,E)$. If $\protect
  \mathcal {G}$ is an expander, then the sampler yields uncorrelated
  configuration with high probability after $t_f=O(\protect \qopname \relax
  o{log}|V|)$ steps; the suggested $t_f$ follows from the loose upper bound
  $|V| \leq m!$. A proof that $\protect \mathcal {G}(L_{\protect \qopname
  \relax m{max}})$ is in fact an expander will depend on the specifics of
  $(\protect \bm {d},\protect \bm {s},L_{\protect \qopname \relax m{max}})$;
  however, we note that $\protect \mathcal {G}(L_{\protect \qopname \relax
  m{max}})$ shares the two fundamental properties with all expanders for
  sufficiently large $L_{\protect \qopname \relax m{max}}$: It is connected and
  not bipartite.}\BibitemShut {Stop}%
\bibitem [{\citenamefont {Newman}(2003)}]{newman2003properties}%
  \BibitemOpen
  \bibfield  {author} {\bibinfo {author} {\bibfnamefont {M.~E.~J.}\
  \bibnamefont {Newman}},\ }\href {\doibase 10.1103/PhysRevE.68.026121}
  {\bibfield  {journal} {\bibinfo  {journal} {Phys. Rev. E}\ }\textbf {\bibinfo
  {volume} {68}},\ \bibinfo {pages} {026121} (\bibinfo {year}
  {2003})}\BibitemShut {NoStop}%
\bibitem [{\citenamefont {Hidalgo}(2016)}]{hidalgo2016disconnected}%
  \BibitemOpen
  \bibfield  {author} {\bibinfo {author} {\bibfnamefont {C.~A.}\ \bibnamefont
  {Hidalgo}},\ }\href {\doibase 10.1007/s41109-016-0010-3} {\bibfield
  {journal} {\bibinfo  {journal} {Appl. Netw. Sci.}\ }\textbf {\bibinfo
  {volume} {1}},\ \bibinfo {pages} {6} (\bibinfo {year} {2016})}\BibitemShut
  {NoStop}%
\bibitem [{\citenamefont {Havel}(1955)}]{havel1955remark}%
  \BibitemOpen
  \bibfield  {author} {\bibinfo {author} {\bibfnamefont {V.}~\bibnamefont
  {Havel}},\ }\href@noop {} {\bibfield  {journal} {\bibinfo  {journal} {Casopis
  Pest. Mat.}\ }\textbf {\bibinfo {volume} {80}},\ \bibinfo {pages} {477}
  (\bibinfo {year} {1955})}\BibitemShut {NoStop}%
\bibitem [{\citenamefont {Hakimi}(1962)}]{hakimi1962realizability}%
  \BibitemOpen
  \bibfield  {author} {\bibinfo {author} {\bibfnamefont {S.~L.}\ \bibnamefont
  {Hakimi}},\ }\href {\doibase 10.1137/0110037} {\bibfield  {journal} {\bibinfo
   {journal} {J. Soc. Ind. Appl. Math.}\ }\textbf {\bibinfo {volume} {10}},\
  \bibinfo {pages} {496} (\bibinfo {year} {1962})}\BibitemShut {NoStop}%
\bibitem [{\citenamefont {Newman}\ and\ \citenamefont
  {Girvan}(2004)}]{newman2004finding}%
  \BibitemOpen
  \bibfield  {author} {\bibinfo {author} {\bibfnamefont {M.~E.~J.}\
  \bibnamefont {Newman}}\ and\ \bibinfo {author} {\bibfnamefont
  {M.}~\bibnamefont {Girvan}},\ }\href {\doibase 10.1103/PhysRevE.69.026113}
  {\bibfield  {journal} {\bibinfo  {journal} {Phys. Rev. E}\ }\textbf {\bibinfo
  {volume} {69}},\ \bibinfo {pages} {026113} (\bibinfo {year}
  {2004})}\BibitemShut {NoStop}%
\end{thebibliography}
\end{document}